\documentclass[aps,twocolumn,pre,showpacs,color,psfig,epsf]{revtex4-1}
\usepackage{amsmath}
\usepackage{bm}
\usepackage{color}
\usepackage{amsfonts}
\usepackage{epsf}
\usepackage{graphicx}
\usepackage{natbib}
\usepackage{textcomp}
\usepackage{tikz}
\usepackage[colorlinks,linkcolor=blue,citecolor=blue]{hyperref}
\definecolor{emil}{RGB}{3,101,3}
\definecolor{francesco}{RGB}{76,76,204}
\newcommand{\fra}[1]{#1}
\newcommand{\emil}[1]{#1}

\newcommand{\C}{\mathcal{C}}
\newcommand{\avg}[1]{\langle#1\rangle}
\newcommand{\Var}{\text{Var }}

\DeclareMathOperator*{\argmax}{arg\,max}

\def\equationautorefname~#1\null{%
	Eq.~(#1)\null
}

\newcommand{\peclet}{P{\'e}clet~}

\begin{document}

\title{Efficiency of one-dimensional active transport conditioned on motility}
%\title{Efficiency of conditioned models of one-dimensional active transport}

\author{F. Cagnetta\textsuperscript{*}, E. Mallmin\textsuperscript{*}}

\affiliation{SUPA, School of Physics and Astronomy, University of Edinburgh, Peter Guthrie Tait Road, Edinburgh EH9 3FD, United Kingdom}

\begin{abstract}
By conditioning a stochastic process on the value of an observable, one obtains a new stochastic process with different properties. We apply this idea in the context of active matter, and condition  interacting self-propelled particles on their individual motility. Using the effective process formalism from dynamical large deviations theory, we derive the interactions that actuate the imposed mobility against jamming interactions in two toy models---the totally asymmetric exclusion process and run-and-tumble particles, \emil{in the case of two or three particles}. We provide a framework which takes into account the energy-consumption required for self-propulsion, and address the question of how energy-efficient the emergent interactions are. Upon conditioning, run-and-tumble particles develop an alignment interaction and achieve a higher gain in efficiency than TASEP particles. A point of diminishing returns in efficiency is reached beyond a certain level of conditioning. With recourse to a general formula for the change in energy efficiency upon conditioning, we conclude that the most significant gains occur when there are large fluctuations in mobility to exploit. From a detailed comparison of the emergent effective interaction in a two- versus a three-body scenario, we discover evidence of a screening effect which suggests that conditioning can produce topological rather than metric interactions.  
\end{abstract}
\maketitle

\section{Introduction}

How should a single biological entity---a macromolecule, a cell, an organism---act to efficiently fulfil its functions in the presence of restrictive collective effects? This question inverts the usual aim of active matter theories, which is to derive the `macroscopic' consequences of a postulated `microscopic' dynamics where fluxes and forces are generated by consuming energy~\cite{ramaswamy2010aa,vicsek2012aa}. \emil{In populations of self-propelled particles, for instance, the efficiency with which the particles convert energy into motion is reduced by collisions~\cite{helbing2001aa}.}
\fra{However, in biological systems equipped with sensing and feedback mechanisms between constituents, we expect well-adapted, or \textit{smart}, interactions to reduce inefficient behaviour like jamming. This suggests that smart interactions of active systems, such as an alignment rule \`a la Vicsek~\cite{Vicsek1995}, might emerge as solutions to physically motivated optimization problems~\cite{Cavagna2014,nemoto2019aa,Tociu2019x}.}

\emil{The idea central to the present work is that smart interactions such as collision avoidance and alignment, can be obtained  by conditioning an active matter model on high values of individual} \fra{motility}. This generalizes to the field of active matter an idea due to R. M. L. Evans~\cite{evans2004aa,RMLEvans2005}, of deriving driven nonequilibrium models with a specific steady-state current from a subensemble of atypical trajectories of an equilibrium process. We summarize \emil{below} how conditioning in this way has been made operational using modern mathematical tools (\ref{sec:Modelmaking}) and how, in this work, we apply it to simple \emil{few-particle} active matter models in \emil{one dimension} (\ref{sec:modelsummary}).

%This idea is central to the present work, which is aimed at obtaining smart interactions (e.g. collision avoidance and alignment) by conditioning \fra{simple one-dimensional active matter models} (e.g. active particles with volume exclusion) on high values of individual \fra{motility}. This generalizes to the field of active matter an idea due to R. M. L. Evans~\cite{evans2004aa,RMLEvans2005}, of deriving driven nonequilibrium models with a specific steady-state current from a subensemble of atypical trajectories of an equilibrium process. We summarize in the Introduction how conditioning in this way has been made operational using modern mathematical tools (\ref{sec:Modelmaking}) and how, in this work, we apply it to simple active matter models (\ref{sec:modelsummary}).

\subsection{Model-making by conditioning}\label{sec:Modelmaking}
\begin{figure}
	\begin{center}
		\begin{tikzpicture}
		\node[anchor=south west,inner sep=0] (image) at (0,0) 				{\includegraphics[angle=-90,width=0.45\textwidth]{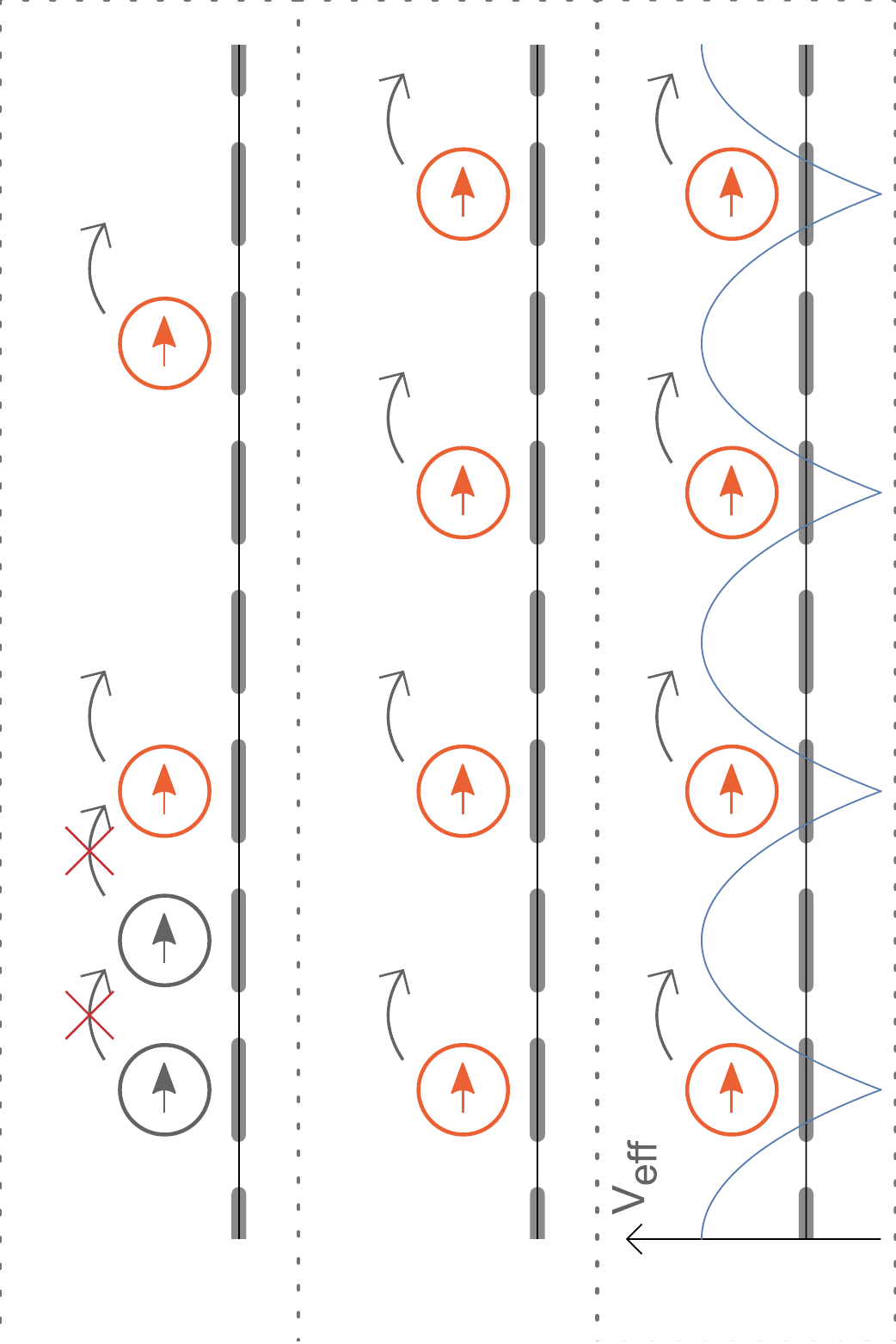}};
		\begin{scope}[x={(image.south east)},y={(image.north west)}]
		\node at (0.03,0.825+0.1) {\bf (a)} ;
		\node at (0.03,0.495+0.1) {\bf (b)} ;
		\node at (0.03,0.165+0.1) {\bf (c)} ;
		\end{scope}
		\end{tikzpicture}		
		\caption{
			Illustration of the effective process generation. \textbf{(a)} Under the naive dynamics, each particle attempts to jump forward at the same rate: some succeed (red), some fail due to steric interactions (grey). \textbf{(b)} Due to the stochasticity of the dynamics, the system can end up in an unlikely state where collective motion has been achieved by mere chance (in the figure because there is some space between particles). \textbf{(c)} The effective process mimics the beneficial fluctuation, by adding to the original process a repulsive interaction between particles.}
		\label{fig:CARTOON}
	\end{center}
\end{figure}
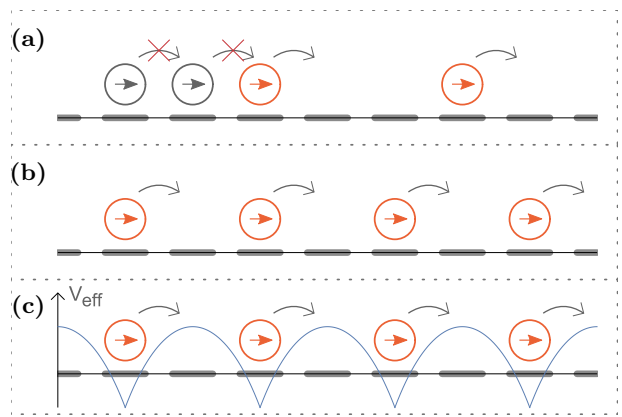

To \textit{condition} a stochastic process is to build a conditioned probability in the classic Kolmogorov sense. If $\Gamma$ is a realization of the stochastic dynamics (a full specification of the trajectories of all constituents) \emil{and $\mathcal{O}(\Gamma) = \mathcal{O}$ denotes a constraint on some trajectory-dependent observable,} the conditioned process is defined via 
\begin{equation}\label{eq:cond}
P(\Gamma\, |\, \mathcal{O}) = \frac{ P(\Gamma\text{ and } \mathcal{O})}{P(\mathcal{O})}.
\end{equation}
\emil{$P(\Gamma\text{ and } \mathcal{O})$ is the probability of observing a specific constraint-fulfilling trajectory $\Gamma$ among all possible trajectories. Dividing by the probability $P(\mathcal{O})$ of realizing the constraint with any consistent trajectory, we obtain a new normalized ensemble $P(\Gamma\, |\, \mathcal{O})$ were every trajectory satisfies the constraint.}
%\fra{where $P(\Gamma\, |\, \mathcal{O}(\Gamma))$ denotes the conditioned probability of a trajectory $\Gamma$ among those realising the constraint $\mathcal{O}(\Gamma)$, $P(\Gamma\text{ and } \mathcal{O}(\Gamma))$ the joint probablity of observing a trajectory $\Gamma$ and satisfying the constraint, $P(\mathcal{O}(\Gamma))$ the probability of satisfying the constraint with whichever trajectory.}
The problem of translating this formal construction into an explicit stochastic dynamics has only recently been solved with some generality. The key assumptions needed are that (1) the observation-time $t$ of trajectories is large compared to the characteristic time-scale(s) of the original dynamics, (2) the \textit{dynamical observable} $\mathcal{O}(\Gamma)$ is time-additive, i.e.\ all its cumulants scale linearly with $t$, and (3) that the original process is Markovian and time-homogeneous. Based on the theory of large deviations, one can extract an \textit{effective process}~\footnote{This process has variously been referred to as \textit{effective, driven, auxiliary} in the literature} whose typical realizations (asymptotically) coincide with the trajectory ensemble implied by the conditioning \eqref{eq:cond} \cite{popkov2010aa,jack2010aa,jack2015aa,chetrite2015conditioning}---i.e.,\ the effective process describes how a \textit{fluctuation}, meaning an atypical value of $\mathcal{O}$, is generated. Remarkably, the effective process is Markovian and time-homogeneous too, and general expressions for its transition rates (discrete case) \cite{jack2010aa} or drift and diffusion functions (continuous case) \cite{chetrite2015conditioning} have been derived. How such a process is generated is illustrated in \autoref{fig:CARTOON}. 

The same effective process emerges from the so-called Maximum Caliber (MaxCal) method \cite{Dixit2018,Monthus2011,Chetrite2015}. Based on the constrained maximisation of a path-wise entropy, MaxCal extends the maximum entropy principle of Jaynes~\cite{Jaynes1957} and yields the effective process when the constraints are made on long-time averages of time-additive observables (non-Markovian ensembles may otherwise result). MaxCal has been applied with success in active matter problems, in order to infer from empirical data the interactions that govern bird flocking~\cite{Bialek2012,Cavagna2014}.

For a Markov process describing a system of interacting particles, the additional effective interactions that emerge upon conditioning can be directly appreciated from the rates of the effective process---if one can find them. That amounts to solving an eigenvalue problem in the dimension of the state space, wherefore analytical results are scarce. There are nonetheless a few notable successes for integrable models, including a range of results for current fluctuations in the TASEP \cite{Derrida1998,derrida2007,prolhac2008,gorissen2012aa,popkov2010aa} and zero-range processes \cite{Harris2013,hirschberg2015}, as well as kinetically constrained models \cite{Garrahan2009aa,Jack2013}. These, together with numerical and analytical evidence from nonequilibrium liquids~\cite{cagnetta2017aa,nemoto2019aa,Tociu2019x,Fodor2019x}, indicate that effective interactions are capable of driving a constrained system towards novel phases. 
There is, however, still a limited understanding of what features of the unconditioned process and conditioning variables lead to interactions that have physical plausibility---firstly, in the sense of their qualitative features, like the range of interactions; secondly, in terms of their energy efficiency. This work is an basic study of this question through a detailed comparison of simple active matter models.
 
\subsection{Description of models and results}\label{sec:modelsummary}

We consider two one-dimensional toy models of active matter: the totally asymmetric exclusion process (TASEP), and interacting run-and-tumble particles (RTPs). In the TASEP on an $L$-periodic lattice, $N$ particles hop clockwise each with a rate $\gamma$, unless the arrival site is already occupied. The RTP model differs only in that each particle has a variable direction $+$ (clockwise) or $-$ (anti-clockwise), alternating or \textit{tumbling} between the two with a rate $\omega$ \cite{Thompson2011}. The TASEP has several active matter interpretations, including motor protein transport and DNA transcription~\cite{Chou2011,chowdhury2005aa}. The RTP dynamic is a simplistic model of microswimmer motility, e.g., the motility patterns of bacteria such as \textit{E. Coli} \cite{Berg2004,schnitzer1993aa}. 
\fra{We refer to the two processes just described as \emph{naive}, }\emil{ to emphasize that these interactions have not been optimized with respect to motility.} 
%so as to distinguish them from the conditioned processes that will be defined later on.

\fra{When interpreted as active agents, both TASEP and RT particles accomplish their function, i.e.\ self-propulsion, with some level of \textit{efficiency}. In the energetic picture we have in mind, a unit of energy is consumed at a rate $\gamma$ (e.g., from the hydrolysis of one ATP molecule \cite{Schnitzer1997}), and is converted into a hop on the lattice if possible; otherwise it is wasted. Therefore, we identify the total number of steps per particle on the lattice as an \emph{output}}. \emil{The efficiency is the ratio of output to input, with the input being the number of energy units consumed}. %  { of the process. The efficiency is then the ratio of this output to the number of energy units consumed or \emph{input}.
 \fra{The output coincides with the particle current for the TASEP, while for the RTPs it is an undirected particle traffic.}

\fra{For both the TASEP and RTPs, exclusion interactions reduce the energy efficiency of motion as defined above.} By conditioning on the number of steps of each particle, we aim to recover the equivalence between energy units consumed and steps taken. However, as clarified below, conditioning alters not only the  interactions between particles, but also the individual base rate of energy consumption. Therefore, the conditioned process is not guaranteed to be more efficient than the naive one, and the efficiency needs to be assessed in both cases and compared.

Due to the lack of general analytical solutions, we proceed numerically but exactly with the conditioning problems, and limit our scope to two and three particles. While the effective interactions have been derived exactly for the $N$-particle TASEP in the limit of a large current \cite{popkov2010aa}, this limit alone is insufficient for our analysis which encompasses also moderate levels of conditioning. In fact, we find that when conditioning the TASEP on higher currents a state of diminishing returns quickly sets in. Further increase in output is accompanied by negligible increase in efficiency. In simpler terms, the main effect of the conditioning is to make the particles jump faster, and at a proportionally higher energy consumption. The effect of the emergent effective interaction---long-range and repulsive \cite{popkov2010aa}---is small in comparison.

The outcome is remarkably different in the RTP model, whose effective process has not been considered elsewhere for more than one particle \cite{Mallmin2019b}. Given a high active \peclet number ($\text{Pe}=\gamma/\omega$), there is a window of fluctuations of the naive process for which the corresponding effective process exhibits directional alignment interactions, with little increase in the base hopping rate. Therefore, the gain in efficiency upon conditioning on higher-than-average motility is substantial. A similar alignment phenomenon was also observed in rare event simulations of active particles~\cite{nemoto2019aa}.

Building on the comparison between the two examples studied, we give a general quantitative argument that a large variance-to-mean ratio in the output, as, for example, afforded by slowly evolving internal states coupled to the output, implies a high attainable gain in efficiency. Furthermore, we present a formal construction of an interaction potential that is guaranteed to increase the efficiency of a naive process whilst keeping the energy consumption fixed. 
Finally, comparing the two- and three-particle conditioned processes allows us to make some concrete statements on open questions regarding the factorization of the effective interactions. For example: when do emergent $N$-body interactions reduce to simpler (e.g., 2-body), and what is the nature of many-body contributions?

\section{Dynamical large deviations formalism}\label{dynLD}
We begin with an overview of the mathematical machinery of dynamical large deviations theory, which allows the explicit construction of the effective process introduced above. We then illustrate its application to interacting particle systems. The theory concerns the asymptotic probability distributions of time-integrated observables of Markov processes 
~\cite{ruelle2004thermodynamic,chetrite2015conditioning,Touchette2018}. A Markov jump process, specifically, is characterised by a vector of probabilities $P$ (with a component for each configuration) evolving by the  master equation $\partial_t P = \mathbb{W}P$. The matrix $\mathbb{W}$ has elements \cite{vanKampen2007}
\begin{equation}\label{eq:naiveproc}
\mathbb{W}_{\mathcal{C'},\mathcal{C}}= W(\mathcal{C}\rightarrow\mathcal{C'}) - \delta_{\mathcal{C'},\mathcal{C}}\sum_{\mathcal{C''}\neq\mathcal{C}}W(\mathcal{C}\rightarrow\mathcal{C''}),
\end{equation}
where $W(\C\to\C')$ denotes the transition rate from configuration $\C$ to $\C'$.
Consider a time-additive observable $N_t(\Gamma)$, e.g., the total number of steps of an active particle for the realization $\Gamma$. To determine the exact time-dependent distribution $P_t(N_t)$ is a daunting task. It is nonetheless often possible to characterise its fluctuations via a \textit{large deviation principle} \cite{ellis2007entropy,touchette2009aa}: 
\begin{equation}\label{eq:largedev}
P(N_t) \asymp e^{-tI(N_t/t)},\quad I(\sigma) = \sup_{s\in \mathbb{R}}\left\lbrace s\sigma - c(s) \right\rbrace.
\end{equation}
The symbol $\asymp$ means equality of the logarithms in the $t\rightarrow\infty$ limit. The \textit{rate function} $I(\sigma)$ is a non-negative function which vanishes at the average $\bar{\sigma} \equiv \lim_{t\to\infty} \avg{N_t}/t$. \emil{For $N_t = \sigma t \neq \bar{\sigma}t$, it gives} \fra{the decay rate of the likelihood of sustaining the fluctuation. The scaled variable $\sigma$ is the effective hopping rate observed over time $t$.} When convex and differentiable, $I(\sigma)$ is the Legendre-Fenchel (LF) transform of the \textit{scaled cumulant generating function} (SCGF) $c(s)$, defined as the long-time limit of $t^{-1}\ln{\left\langle e^{sN_t}\right\rangle}$.

According to the Donsker-Varadhan theory~\cite{donsker1975aa}, $c(s)$ coincides with the principal eigenvalue of the \textit{tilted} transition matrix $\mathbb{W}^{\text{tilt}}(s)$, defined by multiplying each off-diagonal element of $\mathbb{W}$ by $e^{s\alpha(\mathcal{C}\rightarrow\mathcal{C'})}$, where $\alpha$ measures the increase in $N_t$ across the transition $\mathcal{C}\rightarrow\mathcal{C'}$, e.g., 1 if the transition is a hop, else 0. By the Perron-Frobenius theorem \cite{Seneta1981}, $c(s)$ is a real function of $s$.
The spectral elements of $\mathbb{W}^{\text{tilt}}(s)$ also furnish the construction of the effective process---the process whose typical value of $N_t/t$ can be any chosen $\sigma$, and whose typical trajectories coincide with those atypical trajectories generating $\sigma$ as a fluctuation in the original process~\cite{jack2010aa,chetrite2015conditioning}.

In the first step of its construction, the effective process is parametrized by the bias parameter $s$, rather than the desired fluctuation $\sigma$. Its transition rates  $W^{\text{eff}}$ are given by
\fra{\begin{equation}\label{eq:Weff}
\frac{W^{\text{eff}}(\C\to\C',s)}{W(\C\to\C')} =  \frac{\ell(\C',s)}{\ell(\C,s)}\exp\left\{ s \alpha(\C\to\C')\right\},
\end{equation}
where $\ell(s)$ is the left eigenvector or $\mathbb{W}^{\text{tilt}}(s)$ corresponding to the eigenvalue $c(s)$.
The factor $\ell(\C',s)/\ell(\C,s)$ can be} \emil { cast in the form of an `effective' potential difference via the definition}
\begin{equation}\label{eq:Veff}
V(\C,s) \equiv - \log \ell(\C,s).
\end{equation}
The function $\alpha(\C\to\C')$ enters as a \fra{non-conservative} driving force, since it cannot in general be written as a potential difference. In the last step of this construction, any chosen fluctuation $N_t/t = \sigma$ of the original process is made typical in the effective process by substituting for $s$ the \textit{saddle point} value $s(\sigma) = I'(\sigma)$, i.e.\ the maximiser of the LF transform in \autoref{eq:largedev}. This last step requires convexity of $I$ at $\sigma$.

The whole procedure generalizes painlessly when more than one observable is considered, as when we condition a system of $N$ interacting active particles on each particle's output simultaneously. As the observable $N_t$ becomes an $N$-component vector $\mathbf{N}_t$, so do $s$ and $\sigma$, and the product $s\sigma$ in \autoref{eq:largedev} is replaced by a scalar product $\bm{s}\cdot\bm{\sigma}$. However, we will ultimately set the conditional outputs of the different constituents to the same vale $\sigma$, i.e. $\sigma_i =\sigma$, $i=1,\dots,N$. The multidimensional LF transform then becomes
\begin{equation}\label{eq:Nlegendre}
 I(\sigma) = \sup_{\bm{s}\in \mathbb{R}^N}\left\lbrace \left(\sum_{i=1}^N s_i\right)\sigma - c(s_1,\dots,s_N) \right\rbrace,
\end{equation}
where $I(\sigma)$ denotes $I(\sigma_1,\dots,\sigma_N)$ computed at $\sigma_1 = \dots = \sigma_N = \sigma$.
For the limited number of particles considered in this paper, i.e. $N\leq 3$, it is safe to assume due to the particle's indistinguishability that the supremum of \autoref{eq:Nlegendre} is attained on the line $s_1=s_2=\dots=s_N$. In this case, \autoref{eq:Nlegendre} can be replaced with the simpler
\begin{equation}\label{eq:legendre}
 I(\sigma) = \sup_{s\in \mathbb{R}}\left\lbrace N s\sigma - c(s) \right\rbrace,
\end{equation}
where $c(s)$ is a shorthand for $c(s,\dots,s)$.
Although we have verified this assumption \emph{a posteriori} in all the cases examined here, a symmetry breaking for permutations of the particle labelling cannot be excluded in the general case, so that \autoref{eq:Nlegendre} would not reduce to \autoref{eq:legendre}. 

\section{The Two-body conditioning problem}\label{2Body}
\subsection{The Two-Particle TASEP}\label{2TASEP}

We come now to the two-body TASEP conditioning problem. We set the hopping rate $\gamma=1$ without loss of generality by rescaling time. For the TASEP, the efficiency $\eta$ reduces to the ratio of steady state currents of the (effective or naive) interacting and non-interacting processes. Concerning the efficiency of the naive process, we may in fact suppose arbitrary particle number $N$ and (periodic) lattice size $L$. Since all configurations are equally likely in the TASEP steady state, one finds (cf.\ 2.1.1 of \cite{Blythe2007})
\begin{equation}\label{eq:TASEP_naive_eff}
\eta^{\text{TASEP}}_{\text{naive}} = \frac{1 - N/L}{1 - 1/L}. 
\end{equation}

\begin{figure}
	\begin{center}
		\begin{tabular}{cc}
			\includegraphics[angle=-90,width=0.45\textwidth]{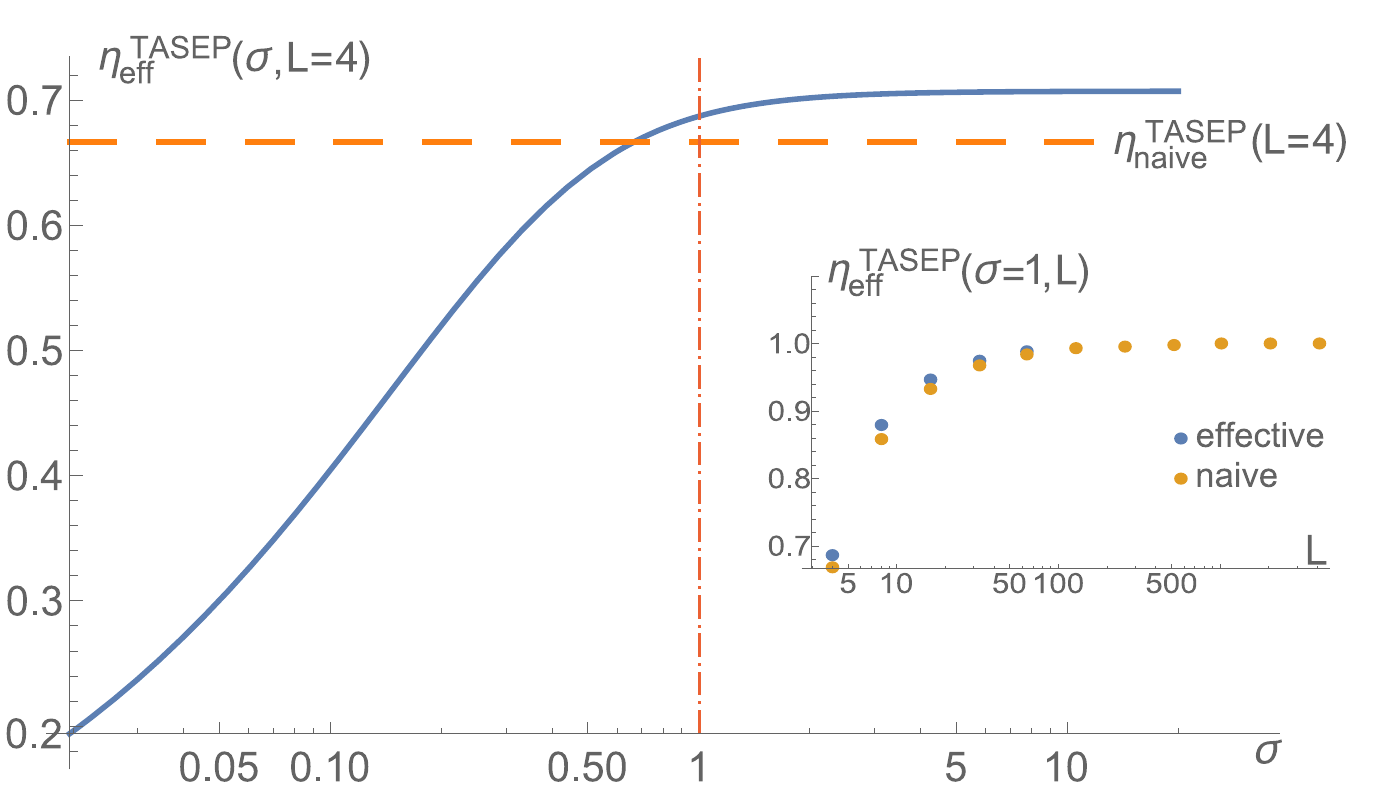}\\
		\end{tabular}
		\caption{
			The efficiency of the two-particle TASEP (blue) can be slightly increased over the naive efficiency ($=2/3$, orange, dashed) by conditioning on larger-than-averge $\sigma$. In the main graph the system size $L=4$ which represents a crowded system. The inset shows how efficiency generally increases with $L$, while the naive--effective difference shrinks; the conditioning variable was chosen to $\sigma=\gamma=1$ (red vertical). }
		\label{fig:TASEPeff}
	\end{center}
\end{figure}

As noted in the introduction, and now demonstrated with reference to \autoref{eq:Weff}, we see that conditioning carries two effects: the addition of an effective interaction potential $V(\C,s)$, and a renormalization of the `bare' hopping rate $1 \to e^{s}$ ($\alpha = 1$ for all allowed transitions). Therefore, the effective-process efficiency for a given level of conditioning $\sigma$ is
\begin{equation}\label{eq:efficiency}
\eta^{\text{TASEP}}_{\text{eff}}(\sigma) \equiv \frac{\sigma}{e^{s(\sigma)}},
\end{equation}
where the dependence on $L$ and $N$ is left implicit. To obtain the saddle-point $s(\sigma)$, we first compute $c(s)$ via a `tilted' Bethe ansatz of the dynamics~\cite{Derrida1998,popkov2010aa}, then solve the maximization \autoref{eq:largedev}. In \autoref{fig:TASEPeff} we plot the resulting efficiency for $N=2$ against $\sigma$ and $L$. The efficiency gain with respect to the naive process is small, and rapidly diminishes with larger system size $L$ (see inset). Just as in the analytically tractable case of large current fluctuations, the effective interaction for moderate conditioning is still a weak long-range repulsion. It is `smart' in the sense of reducing the tendency to jam, but it does not contribute substantially to the hopping rate (at $\sigma=1$, $e^{\Delta V} \lesssim 1.03$ for $L=16$ and decreases with $L$). Rephrasing, the most probable way for the two particles to be as active as in the absence of crowding (i.e., choosing $\sigma=\gamma$) is to simply `push harder'. As this requires more energy input, the efficiency quickly reaches a point of diminishing returns.

\subsection{Two Run-And-Tumble Particles}\label{2RTP}

\begin{figure}
	\begin{center}
		\begin{tikzpicture}
		\node[anchor=south west,inner sep=0] (image) at (0,0)
		{\includegraphics[width=0.45\textwidth]{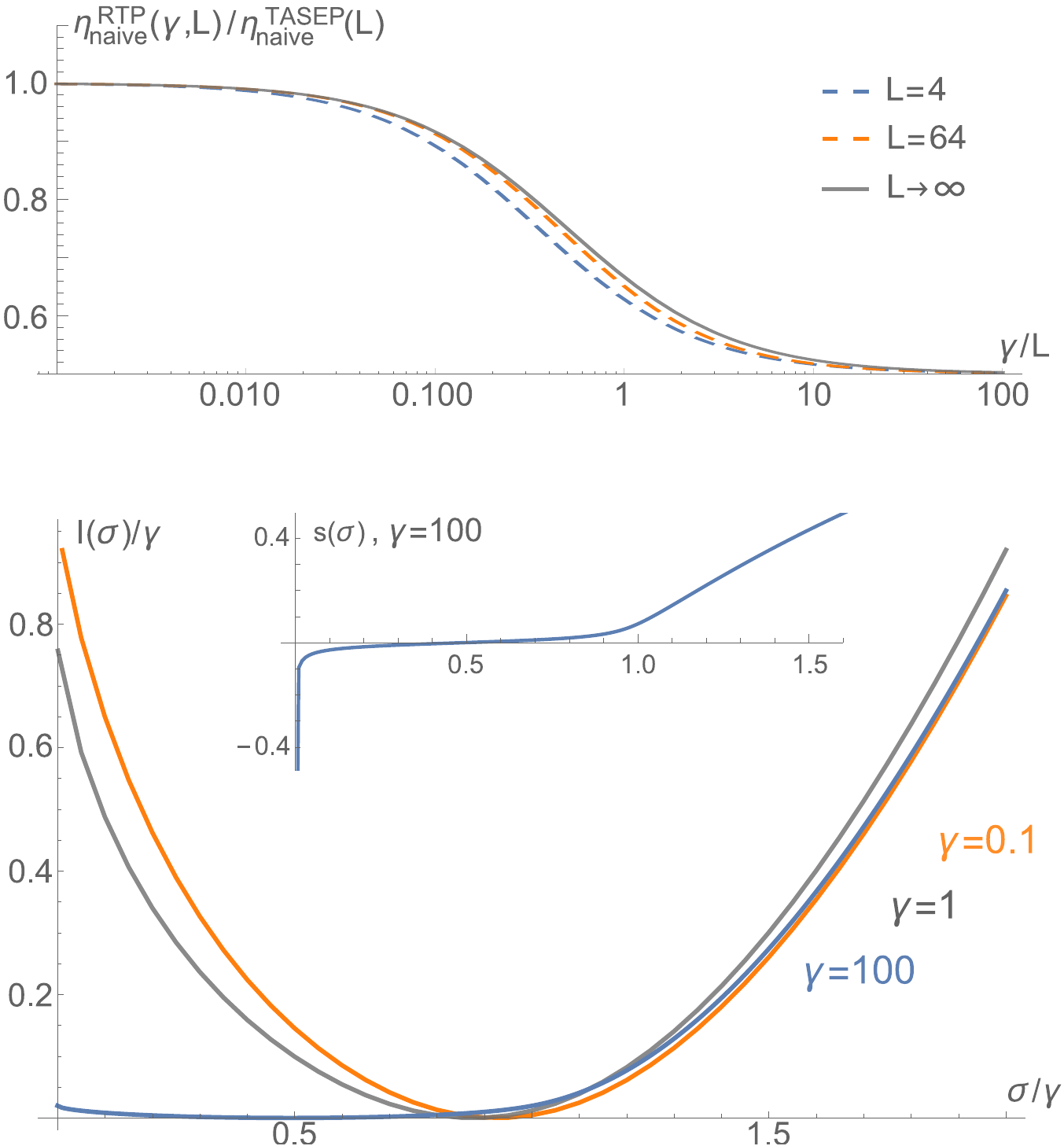}};
		\begin{scope}[x={(image.south east)},y={(image.north west)}]
		\node at (0.01,0.98) {\bf (a)} ;
		\node at (0.01,0.54) {\bf (b)} ;
		\end{scope}
		\end{tikzpicture}
		
		\caption{ 
			Two naively interacting RTPs. \textbf{(a)} The efficiency (from exact formula, not displayed) approaches a scaling form as the system size $L$ becomes large. As the \peclet number becomes comparable to system size, the efficiency drops significantly due to jamming. \textbf{(b)} The rate function \autoref{eq:largedev} for the total number of hops develops a flat region for large \peclet numbers. This region extends beyond the average $\bar{\sigma}$ up to about $\sigma \approx\gamma$. Consequently, the saddle point $s(\sigma) = I'(\sigma)$, shown in the inset for $\gamma = 100$, is close to zero in this range. All graphs show $L=32$; $I(\sigma)$ was computed by LF transform of $c(s)$, in turn obtained by a numerical diagonalization of the tilted Markov matrix. }
		\label{fig:LargeDev}
	\end{center}
\end{figure}

The conclusions are substantially different for the RTP model, which we now consider for $N=2$. Upon rescaling time so as to set the tumbling rate $\omega=1$, the hopping rate $\gamma$ can be interpreted as the active \peclet number $\text{Pe }= \gamma/\omega$, which quantifies the ratio of self-propulsion to diffusion. At any given time, the directions $\tau_i\in \{+,-\}$, $i=1,2$, of the particles may be either aligned or anti-aligned---crucially, a pair of particles may be found in a jammed configuration where each obstructs the other. The exact nonequilibrium steady state of the RTP model is only known for $N=2$ \cite{slowman2016aa}; there, the jammed configuration carries an anomalously large weight. From this solution we obtain an explicit expression for the efficiency of the naive two-particle process. In particular, it has a simple scaling form for large $L$,
\begin{equation}\label{eq:largeLEffRTP}
 \eta^{\text{RTP}}_{\text{naive}}(\gamma) \simeq \frac{1+\gamma/L}{1+2\gamma/L}\quad (N=2).
\end{equation}
For smaller $L$, as shown in the top panel of \autoref{fig:LargeDev}, the exact efficiency curve collapses approximately onto the scaling form \autoref{eq:largeLEffRTP}, provided the RTP efficiency is normalised by the $L$-dependent TASEP efficiency, \autoref{eq:TASEP_naive_eff}. As one would expect, the efficiency drops with increasing \peclet number: when $\gamma \gg L$, the particles will with equal probability be either jammed or in an aligned TASEP configuration, thus mustering only half the TASEP efficiency on average.

\begin{figure}
	\begin{center}
		\begin{tabular}{cc}
			\includegraphics[angle=-90,width=0.45\textwidth]{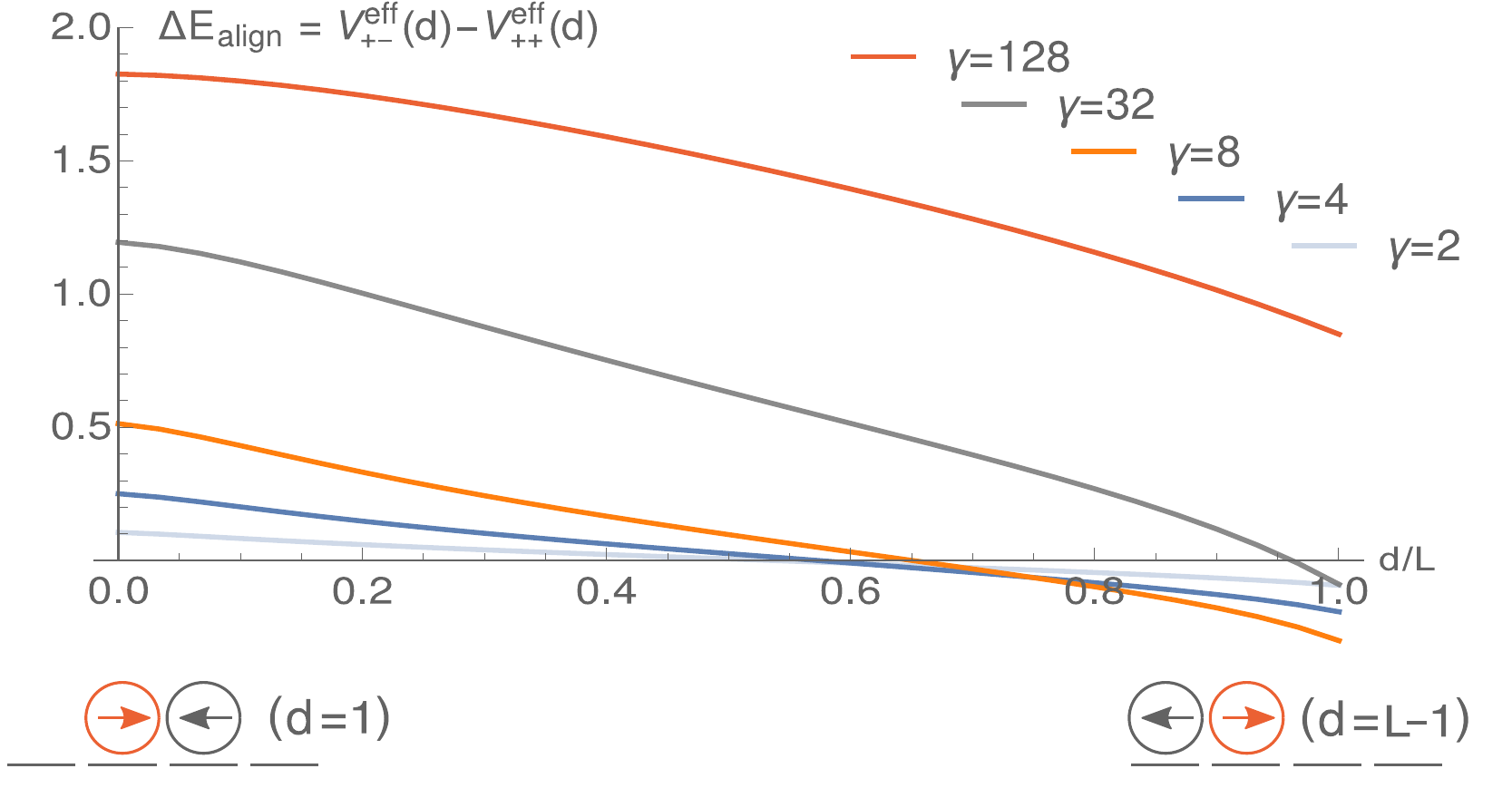}\\
		\end{tabular}
		\caption{This figure illustrates the effective interaction that emerges upon conditioning two RTPs on the value $\sigma = \gamma$; we keep $L=32$ and vary $\gamma$. %---the figure changes only minutely if $L$ and $\gamma$ are similarly rescaled. 
		The tendency to align is largest at short face-to-face distance, as quantified by $\Delta E_{\text{align}}(d)$, where $d$ is the distance from particle one (red) to two (black). When $\gamma \lesssim L$ and the particles are close and back-to-back, this configuration is preferred as the particles have a chance of increasing their distance and then aligning; if $\gamma \gg L$, anti-alignment quickly leads to jamming, and therefore aligning is at all separations the most likely way to increase mobility. The potential \autoref{eq:Veff} was found from a numerical calculation of the spectral elements of the tilted transition matrix.}
		\label{fig:RTPVeff}
	\end{center}
\end{figure}
Next, we construct the effective process and determine its efficiency. Consider first the large deviations of the total number of steps $N_t$ per particle. As shown in bottom panel \autoref{fig:LargeDev}, the naive process average $\bar{\sigma} = \avg{N_t}/t$ (i.e., the zero of $I(\sigma)$) decreases relative to $\gamma$ as this parameter becomes large. The SCGF can be calculated numerically either directly from the tilted transition matrix or by solving a tilted version of the `root-paramterized eigenvalue equations' derived in \cite{Mallmin2019a}. The resulting rate function has a Gaussian profile for fluctuations larger than $\sigma\simeq \gamma$, whereas the complementary regime of fluctuations smaller than the free-particle speed becomes almost flat for large \peclet number. This large variance stems from the particles' ability to either align and produce a large current, or anti-align and then quickly reach the jammed state~\cite{cagnetta2017aa}. This feature proves instrumental in increasing the efficiency of the RTP process. As the inset of \autoref{fig:LargeDev} shows, in the approximate window $\sigma \in [\bar{\sigma}, \gamma]$ where the rate function is flat, the saddle point (which does depends on $\gamma$ for the RTPs) $s_\gamma(\sigma)=I'(\sigma)$ is close to zero. Conditioning the process on $\sigma$ is this range will, beyond the potential \autoref{eq:Veff}, only weakly alter the hopping rates, as $e^{s\alpha}\approx 1$. In this way, the output can be increased without immediately encountering diminishing returns.

For each orientation sector $(\tau_1,\tau_2)$ we get via \autoref{eq:Weff} the $\sigma$-dependent effective potential $V_{\tau_1\tau_2}(x_2-x_1)$ (with $V_{++} = V_{--}$ and $V_{+-}(d) = V_{-+}(L-d)$ by symmetry). The largest and most relevant potential difference is the alignment affinity  $\Delta E_{\text{align}}(d) \equiv V_{+-}(d)-V_{++}(d)$ shown in \autoref{fig:RTPVeff}. The alignment interaction is strongest at short face-to-face distance, and is superimposed on a weak long-range repulsion similar to that of the TASEP for large $L$ and/or small $\gamma$. In addition, \autoref{fig:RTPVeff} suggests that stronger interactions emerge, together with the `flat' branch of the rate function, when $\gamma$ exceeds the ring size. The present result for two particles should then be relevant for systems with many interacting particles where the typical inter-particle distance replaces the ring size---the features of the large deviations~\cite{cagnetta2017aa,nemoto2019aa} do not seem to vary qualitatively by this generalisation.

\begin{figure}
\begin{center}
  \begin{tabular}{cc}
       \includegraphics[angle=-90,width=0.45\textwidth]{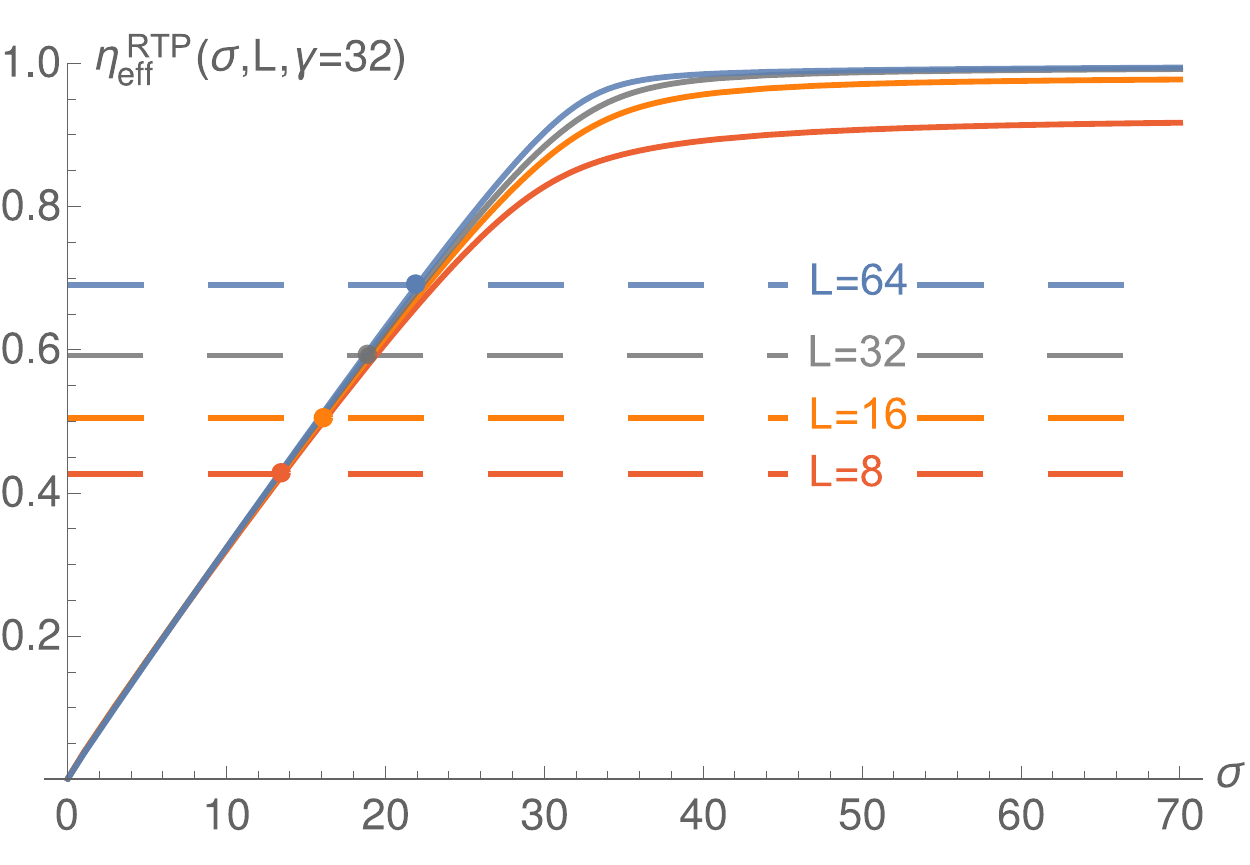}\\
  \end{tabular}
\caption{
	The efficiency of two interacting RTPs conditioned on higher-than-average $\sigma$  (solid lines) increases significantly over the naive efficiency (dashed lines), especially at large $\gamma/L$. The rate of increase in efficiency at the intersection with the naive value (marked by a dot) is given by \eqref{eq:etaprime}.  }
\label{fig:RTPeff}
\end{center}
\end{figure}

The efficiency depends separately on $\gamma$ and $\sigma$ (since the saddle point does) as
\begin{equation}\label{eq:effRTP}
\eta^{\text{RTP}}_{\text{eff}}(\sigma,\gamma) = \frac{\sigma}{\gamma} e^{-s_\gamma(\sigma)},
\end{equation}
which we plot in \autoref{fig:RTPeff} versus $\sigma$. As anticipated, most of the possible gain in efficiency occurs before $\sigma\approx \gamma$, i.e., with little jump rate renormalization, after which diminishing returns sets in and the efficiency plateaus.

%\section{The Three-body conditioning problem (and beyond)}\label{3bodies}
\section{Beyond the two-body conditioning problem}\label{3bodies}

\autoref{eq:effRTP} is not limited to active transport problems. Its formulation presupposes a collection of $N$ entities, each independently receiving an input quantity at a rate $\gamma$. In a (Markovian) collective process, this quantity is converted into an output $\sigma$ (per entity) that obeys a dynamical large deviation principle. We first explore the general implications of this setting. Then we will specialize on the three-body TASEP and RTP problems.

The derivative of $\eta$ with respect to $\sigma$, evaluated at the naive average $\bar{\sigma}$, 
quantifies the immediate improvement in efficiency upon conditioning:
%
%Here we extend the results of \autoref{2Body} to systems of three particles. \emil{Before proceeding with the detailed $N=3$ analysis, let us predict the change in efficiency upon conditioning on the output under much more general conditions. The starting point is \autoref{eq:effRTP}, which is not limited to active transport problems. Its formulation presupposes a collection of $N$ entities, each independently receiving an input quantity at a rate $\gamma$. In a (Markovian) collective process, this quantity is converted into an output $\sigma$ (per entity) that obeys a dynamical large deviation principle. The derivative of $\eta$ with respect to $\sigma$, evaluated at the naive average $\bar{\sigma}$, quantifies the immediate improvement in efficiency upon conditioning:
\begin{equation}\label{eq:etaprime}
\eta'(\bar{\sigma})
= \frac{1}{\gamma}\left[ 1-\frac{\bar{\sigma}}{N\left(\overline{\sigma^2} -\bar{\sigma}^2\right)	}\right],
\end{equation}
where we have used the general identities $s(\bar{\sigma}) = 0$, $N s'(\bar{\sigma}) = I''(\bar{\sigma}) = 1/\text{Var }\sigma$ \footnote{$I''(\bar{\sigma})$ gives the reciprocal variance, as fluctuations close to the mean are Gaussian by the central limit theorem.}, with $N$ the number of constituents. Note that since $\sigma$ is defined \textit{per entity}, it scales as $1/N$. Therefore the subtracting term in \autoref{eq:etaprime} is not ensured to vanish in the large-$N$ limit. If the input-to-output conversion follows strictly Poisson statistics (as for non-interacting particles on a lattice), $\eta'(\bar{\sigma}) = 0$.

\emil{The general conclusion afforded by \autoref{eq:etaprime} is that a large variance-to-mean ratio in the output implies high possible gains in efficiency by conditioning. In effect, when there is ample variance in output, conditioning may produce a more optimized process by chiefly retaining the high-perfomance trajectories of the original process and discarding low-performance ones. Consider again RTPs at high \peclet number, as in the above numerical study for $N=2$. The large variance in mobility is afforded by the separation of time-scales between the re-orientation and hopping events. We therefore expect that the efficiency of self-propelled particle systems can be increased by exploiting fluctuating internal states coupled to the current-generating dynamics~\cite{Pietzonka2016,Mallmin2019b}. As \autoref{eq:etaprime} holds also for large $N$, it could be directly applied to active models for which the large deviation functions have been determined from simulation or by other means, e.g. \cite{nemoto2019aa}.}

%\emil{A modification of the derivation of bounds on fluctuations in \cite{Gingrich2016} shows that $\eta'(\sigma) \geq 0$ for the TASEP, so that conditioning is guaranteed to be beneficial. The RTP lower bound, instead, is negative.}
%\fra{[EMIL-REVISE]}By a simple variation of an established approach to derive bounds on fluctuations \cite{Gingrich2016}, we show in \autoref{sec:bounds} that $\Var \sigma \geq \bar{\sigma}^2 / (\bar{\sigma} + \bar{r})$ where $r$ is the the number of transitions (per unit time) different that are not hops. Then $\gamma \eta'(\bar{\sigma}) \geq - \bar{r}/\bar{\sigma}$. For the TASEP, $r = 0$, so that $\gamma \eta'(\bar{\sigma}) \geq 0$, i.e., conditioning can never decrease efficiency, even if the gains, as demonstrated, may be modest. For RTPs, $r$ refers to tumbles. If we have sufficient time-scale separation, $\gamma \eta'(\bar{\sigma}) \gtrsim 0$.\fra{[]}

\begin{figure}
	\begin{center}
		\begin{tabular}{cc}
			\includegraphics[angle=-90,width=0.45\textwidth]{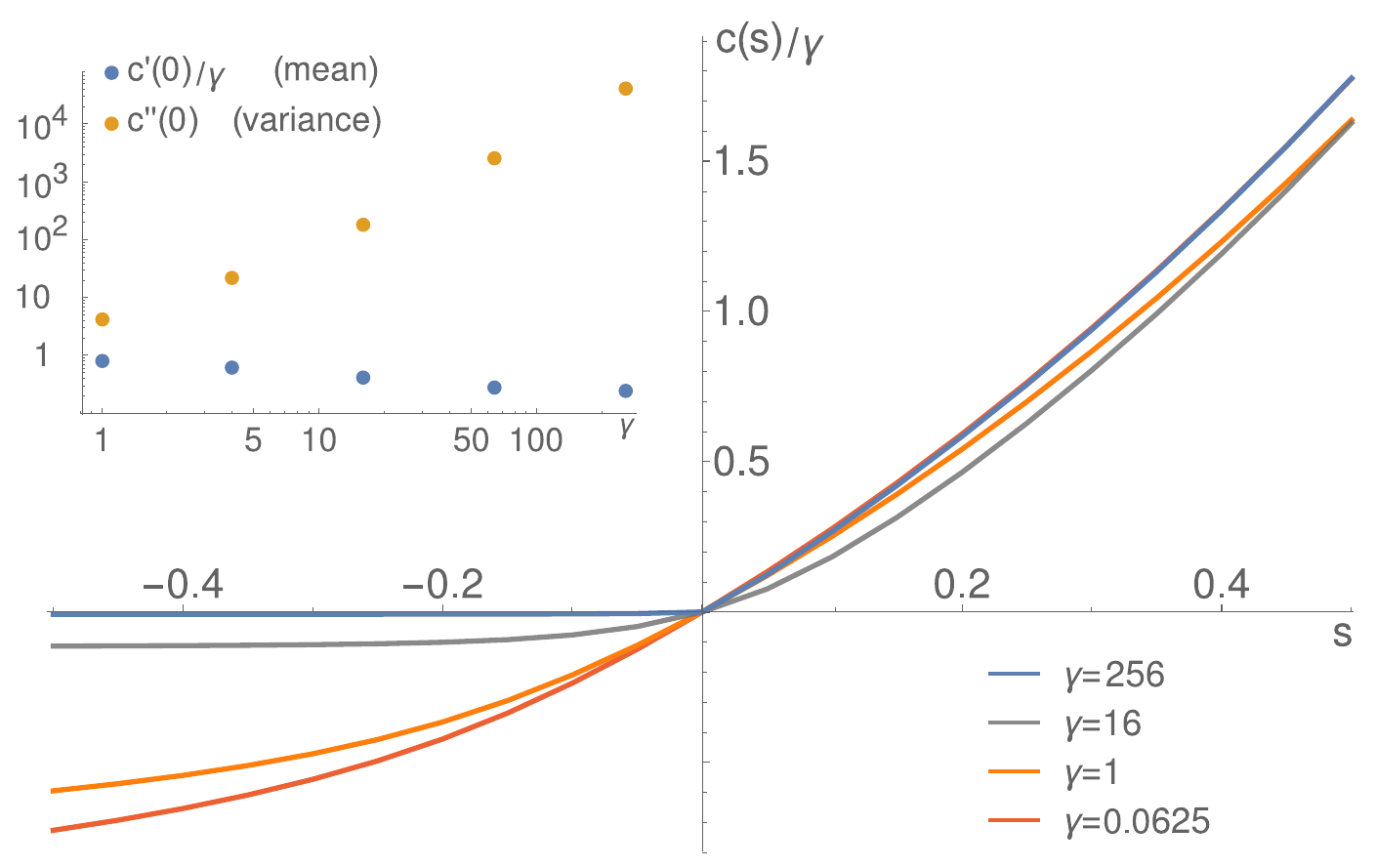}\\
		\end{tabular}
		\caption{SCGF $c(s)$ scaled by $\gamma$ for $N=3$ RTPs, with $\gamma$ in the key. Notice the approximate overlap for $s>0$, analogous to that observed for $N=2$ and shown in \autoref{fig:LargeDev}. The inset shows the mean (blue dots) and variance (yellow dots) of $\sigma$, computed from the $c(s)$ derivatives, as functions of $\gamma$.}
		\label{fig:3RTPSCGF}
	\end{center}
\end{figure}

We now put three particles on the lattice, and study numerically the same conditioning problem as for two RTPs or TASEP-particles in \autoref{2Body}. The lattice size is fixed to $L=16$. The apposite questions to ask are, firstly, whether the observation for two particles (viz., the alignment interaction for the RTPs) generalizes to higher particle numbers; secondly, if the effective three-body potentials are the sum of the pairwise potentials obtained from the two-body conditioning.
Regarding the first question, \autoref{fig:3RTPSCGF} shows the SCGF $c(s)$ for three RTPs, for a range of $\gamma$. By scaling $c(s)$ with $\gamma$, we achieve an approximate superposition of the curves in the $s>0$ half-plane. Therefore, according to \autoref{eq:legendre}, the rate functions superimpose for $\sigma>\bar{\sigma}$, as they do for $N=2$ (cf. \autoref{fig:LargeDev}). Additionally, the second derivative at $s=0$, $c''(0)$ (\autoref{fig:3RTPSCGF}, inset), increases quite steeply with $\gamma$. By Legendre duality, the rate functions will become progressively flatter, as it does for $N=2$.
The peculiar large deviations of interacting RTPs are then preserved in the passage from $N=2$ to $N=3$. Furthermore, we can use \autoref{eq:etaprime} to predict the expected efficiency gain. The inset of \autoref{fig:3RTPSCGF} shows both the mean $\bar{\sigma}$ and the variance $\Var{\sigma}$. Their ratio is already in the hundreds for $\gamma=16$, 
indicating an efficiency derivative close to the upper bound $1/\gamma$.

\begin{figure}
	\begin{center}
		\begin{tabular}{cc}
			\includegraphics[angle=-90,width=0.45\textwidth]{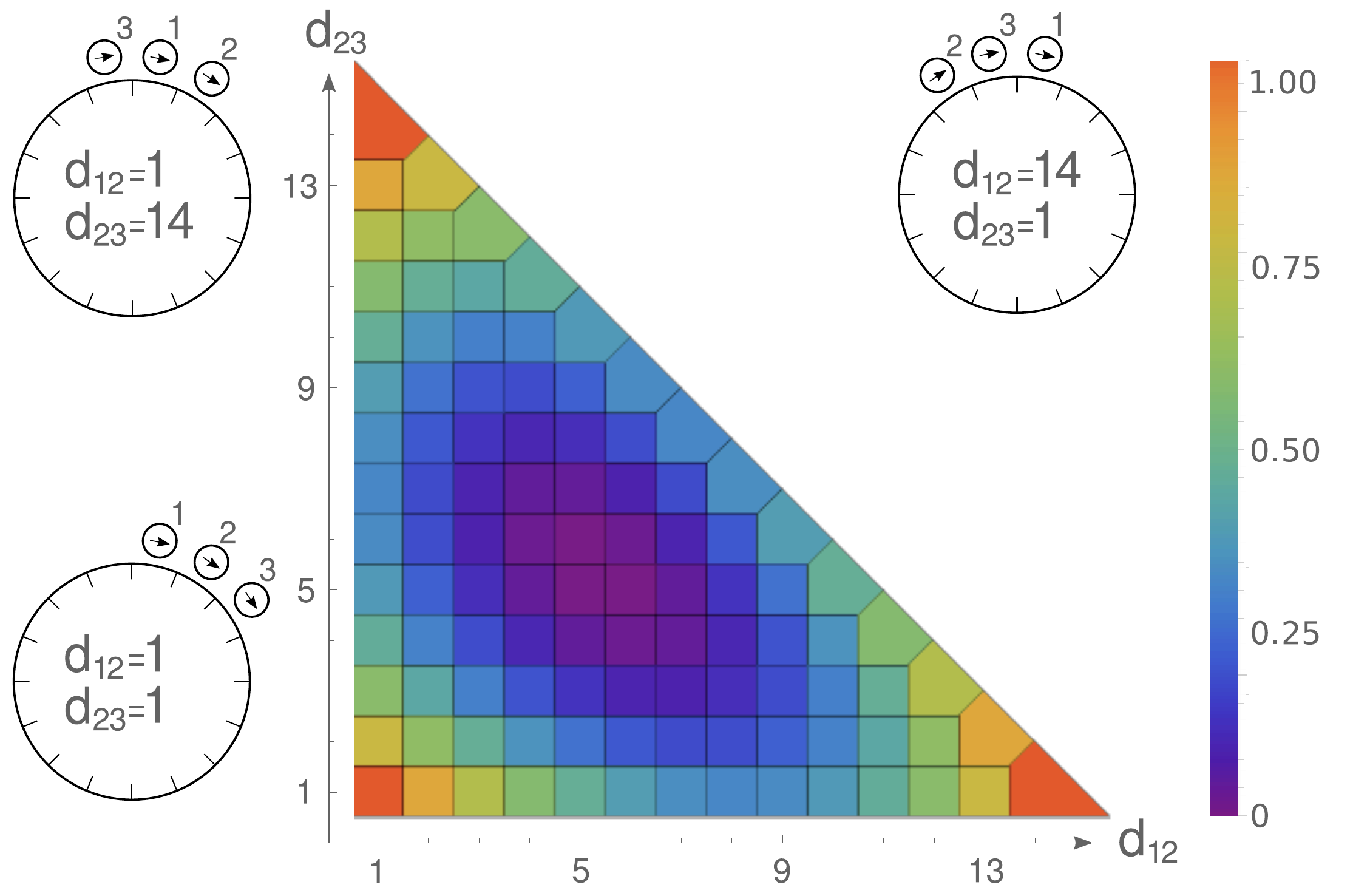}\\
		\end{tabular}
		\caption{
			Effective potential of the three-particle TASEP, with $L=16$ and $s=0.15$, as a function of $d_{12}$ and $d_{23}$. The potential, whose value is represented by the colour, is measured with respect to the minimum at the maximum-distance configurations ($d_{12},d_{23}= 5,5$, $6,5$ and $5,6$, at the centre of the colour plot). The snapshots on the left-hand side and top-right corner depict the three maximum-potential configurations where the particles are next to each other, corresponding to the corners of the colour plot.}
		\label{fig:3PotPic}
	\end{center}
\end{figure}
The TASEP large deviations also do not change appreciably from $N=2$ to $N=3$, especially for large $L$.  Nevertheless, it is interesting to compare the effective potentials obtained in the two cases.
For the potential to be pairwise, the total potential of each $N=3$ configuration must coincide with the sum of the $N=2$ potentials of the three particle pairs. A conceptually important issue immediately arises of what respective levels of conditioning make two- and three-body systems meaningfully comparable. It may at first seem physically intuitive to compare the $N=2$ and $N=3$ systems for the same output per particle $\sigma$.
However, although the large deviations are qualitatively similar, there is a quantitative dependence on the particle density such that, for instance, $\bar{\sigma}_{N=2}>\bar{\sigma}_{N=3}$, especially for small $L$. Importantly, the saddle-point function $s(\sigma)$ is $N$-dependent, giving differing renormalizations $e^{s(\sigma)}$ of the base hopping rates in the $N=2$ and $N=3$ processes conditioned on the same $\sigma$. This suggests that processes with different particle numbers should be compared at fixed $s$ rather than $\sigma$. In addition, as the effective potential is determined by the left eigenvector $\ell(\C,s)$, it is likely to have a simpler algebraic dependence on $s$ than it does on $\sigma$ via $s(\sigma)=I'(\sigma)$. In fact, in the one and only known case where the effective potential factorises, it does so as function of $s$. Nonetheless, the clearest results will be found in the limits of small/large $s$, equivalent to the small/large $\sigma-\bar{\sigma}$ limits.

\subsection{TASEP three-body potential}\label{3TASEP}

\begin{figure}
	\begin{center}
		\begin{tabular}{cc}
			\includegraphics[angle=-90,width=0.45\textwidth]{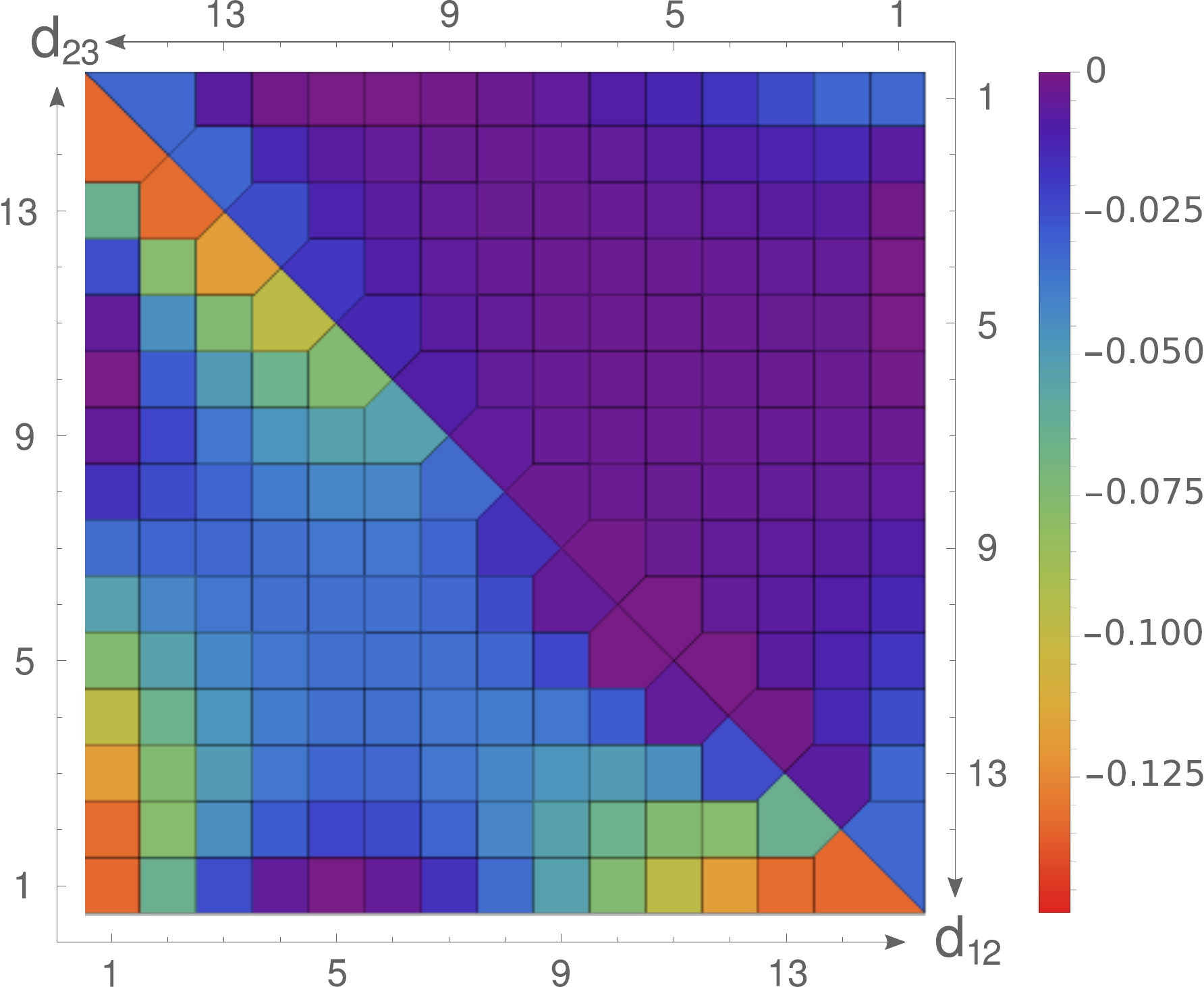}\\
		\end{tabular}
		\caption{
			Fixed-$s$ potential difference $\Delta=V^{(3)}-\widetilde{V}^{(3)}$ between the effective three-body potential and sum the of effective pairwise potentials from the two-particle conditioning\fra{, for the TASEP}. In the upper colour plot $s=2.5$ (high $s$), and in the lower $s=0.15$ (low $s$).}
		\label{fig:TASEP-HiLo}
	\end{center}
\end{figure}

Let us then consider the effective potential  $V^{(3)}(d_{12},d_{23})$ of the three-particle TASEP, where $d_{ij}$ is the distance (in number of lattice sites) from particle $i$ to $j$, with periodicity demanding $d_{31} = L - d_{12}-d_{23}$. As in the $N=2$ problem, the potential is generally repulsive, with maxima at $d_{12},d_{23} = 1, L-2$. This is clearly manifest in \autoref{fig:3PotPic}.
We now compare, for $s$ and $L$ fixed, the potential $V^{(3)}$ to the pairwise potential $\widetilde V^{(3)}$ constructed as the sum of the effective 2-body potentials $V^{(2)}$ found in the previous section; $\widetilde{V}^{(3)}(d_{12},d_{23}) = V^{(2)}(d_{12}) + V^{(2)}(d_{12}) + V^{(2)}(L-d_{12}-d_{23})$. The difference $\Delta = V^{(3)} - \widetilde{V}^{(3)}$ indicates the extent to which the 3-body interaction is reducible to pairwise interactions---which we refer to as \textit{factorization} (of the left principal eigenvector of $\mathbb{W}^{\text{tilt}}_s$).
\autoref{fig:TASEP-HiLo} shows $\Delta$ as a function of $d_{12}$ and $d_{23}$, for $L=16$ and two values of $s$. The top-right colour plot refers to the high-$s$ case, the bottom-left to the low-$s$ case considered also in \autoref{fig:3PotPic}. For the larger $s$, $\Delta \simeq 0$, following the large-$s$ factorisation of the TASEP effective potential, demonstrated in~\cite{popkov2010aa}. As $s$ is reduced, $\Delta$ decreases, signalling that three-body interactions play a significant role in optimally achieving the fluctuation.

We find this reduction to be more pronounced when the three particles are all next to each other (corners of \autoref{fig:TASEP-HiLo}). Our result indicates the cost in (effective) potential of keeping the three particles from colliding to be less than that of keeping the three particle pairs singularly disjoint. This is caused by a sort of screening effect due to the third particle. If, e.g., $d_{12}=d_{23}=1$ (as in the bottom-left corner of \autoref{fig:3PotPic}), then the repulsion between $1$ and $3$ is already effected by the repulsion between 1--2 and 2--3. Alternatively, the result can be resolved with an interaction whose strength decreases not only with the \emph{metric} distance, i.e. the length in lattice units of the shortest path between two particles, but also with the \emph{topological} distance, that is the number of other particles located on this shortest path~\cite{Ballerini2008aa}.

The difference $\Delta$ also reveals the directional asymmetry of $V^{(3)}$,
a purely three-body effect caused by the unidirectional motion of TASEP. Imagine, for instance, fixing the first two particles and moving the third along the lattice, thus exploring the $d_{12}=1$ vertical line of the potential landscape shown in \autoref{fig:3PotPic}. In moving from $d_{23}=1$ to $L-2$ (the maximum distance for a given $L$), the pairwise potential $V^{(2)}$ reaches its minimum at the midpoint and is symmetrical. The minimum of $V^{(3)}$, instead, is slightly shifted towards the $d_{23}=L-2$ end. In simple terms, $V^{(3)}$ favours configurations where particle $3$ lies behind the small cluster formed by $1$ and $2$, rather than that where $d_{23}=d_{31}$. This effect, which resembles slipstreaming (in the absence of any fluid), is clearer in the difference $\Delta$ (cf. \autoref{fig:TASEP-HiLo}) than in the three-body potential itself.

\subsection{RTP three-body potential}\label{3RTP}

We now discuss the three-body effective potential for a system of RTPs. The effective potential depends on the orientational as well as translational degrees of freedom. We then write the potential as $V_{\tau_1\tau_2\tau_3}^{(3)}(d_{12},d_{23})$, where $d_{ij}$ is the distance from particle $i$ to $j$ and $\tau_i \in\{ +,-\}$ is the orientation of the $i$th particle. For the TASEP(\autoref{3TASEP}), we only had the single orientation sector $\tau_1\tau_2\tau_3={{+}{+}{+}}$; for the RTPs, we consider only the sectors ${+}{+}{+}$ and ${-}{+}{+}$, as the rest are related to these by permutation of particle labelling and spatial inversion.

\begin{figure*}
	\begin{center}
		\begin{tabular}{cc}
			%{\centering \hspace{0cm}$s=0.15$ \hspace{7cm} $s=2.5$}\\
			\includegraphics[angle=-90,width=0.45\textwidth]{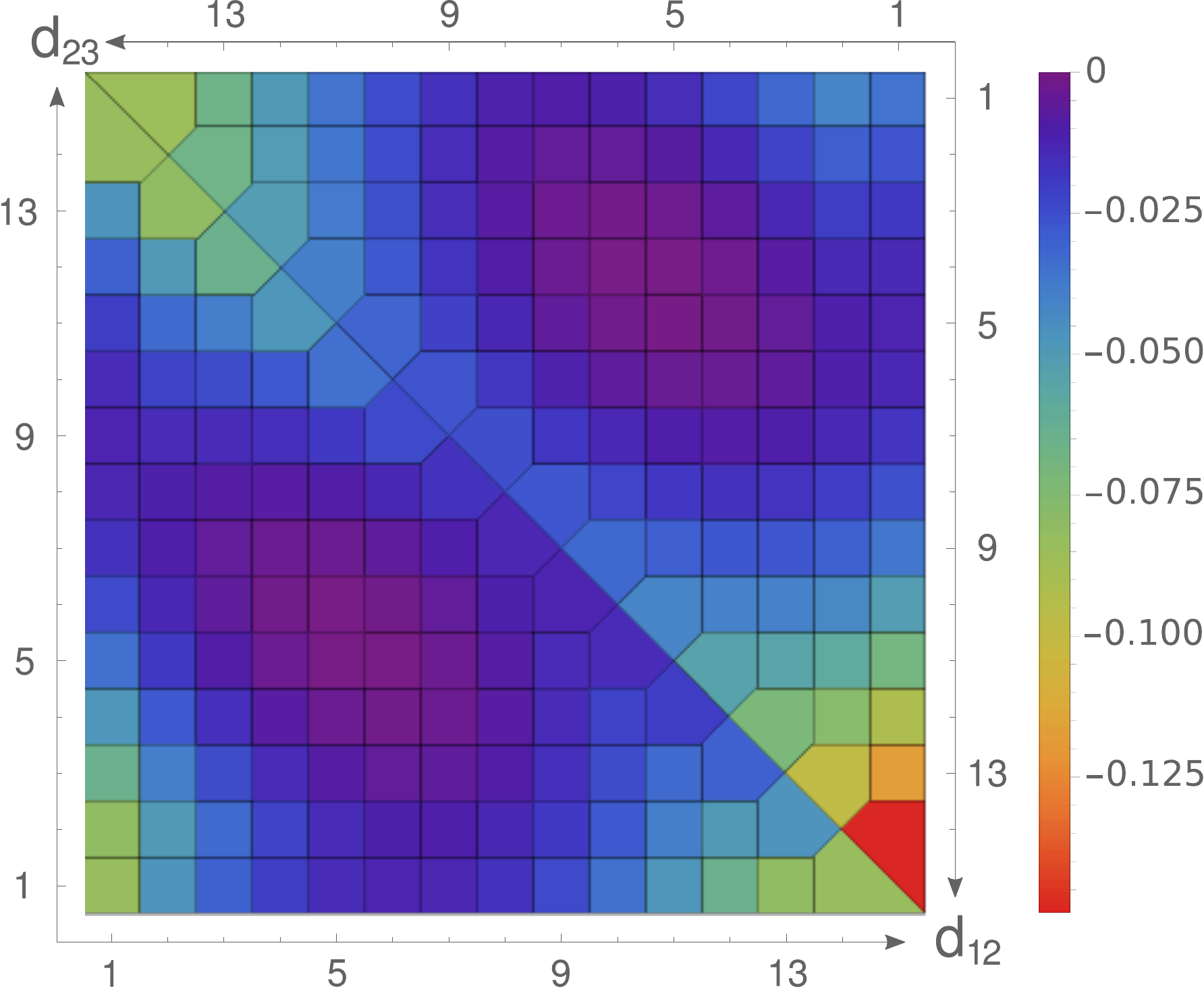}
			\includegraphics[angle=-90,width=0.45\textwidth]{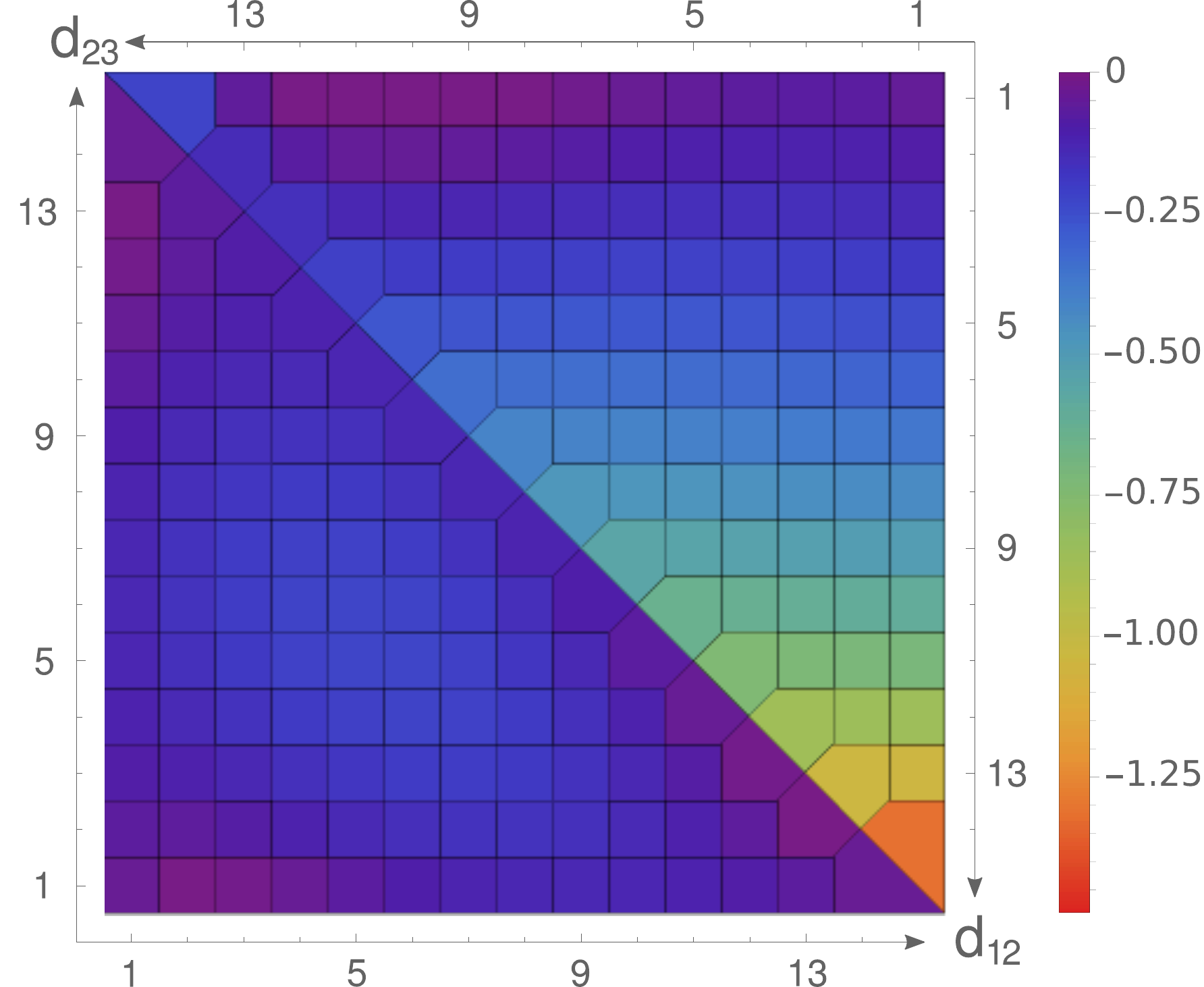}		
		\end{tabular}\\[8pt]
		{\hspace{0cm} \textbf{(a)} $s=0.15$ \hspace{6.3cm} \textbf{(b)} $s=2.5$}
		\caption{
			Plot of the discrepancy $\Delta = V^{(3)} - \widetilde{V}^{(3)}$ between three-body potential and sum of pairwise interactions for the RTP, for \textbf{(a)} low $s$-value and \textbf{(b)} high $s$-value. Lower triangles show $\Delta_{+++}$, and upper triangles $\Delta_{-++}$. Parameters are $L=16$ and $\gamma=1$.}
		\label{fig:RTP-HiLo}
	\end{center}
\end{figure*}
Plotting $V_{+++}^{(3)}(d_{12},d_{23})$ would reveal the same weak repulsive interaction found for the 3-particle TASEP and shown in \autoref{fig:3PotPic}; the difference $V_{+++}^{(3)}-V_{-++}^{(3)}$ shows an alignment interaction reminiscent of that discussed in \autoref{2RTP}. Additional three-body contributions are better understood by resorting to the difference with respect to the sum of pairwise interactions, $\Delta_{\tau_1\tau_2\tau_3}$---notice the dependence on particle orientations for RTPs. This difference is shown for both the ${+}{+}{+}$ (bottom-left triangle) and ${-}{+}{+}$ (top-right triangle) sectors in \autoref{fig:RTP-HiLo}, at $\gamma=\omega=1$ and $L=16$. The left and right panels are representative of the low $s$ and high $s$ regimes, respectively. Let us begin with the former, i.e. compare $\Delta_{+++}$ and $\Delta_{-++}$ at low $s$.

$\Delta_{+++}$, though generally small, is larger in modulus at the corners of the colour plot, implying a weaker repulsion than in the two-body case. This observation, as in the TASEP, can be explained by a screening effect due to the third particle. $\Delta_{-++}$ displays a similar landscape, apart from two differences. First, the well at $d_{12}=L-2, d_{23}=1$ is deeper than for $\Delta_{+++}$ (see \autoref{fig:RTP-HiLo}, left panel, bottom-right corner). This is a jammed configuration, which is obtained by the $d_{12}=L-2, d_{23}=1$ configuration shown in the top-right corner of \autoref{fig:3PotPic} by flipping the arrow of particle $1$: the rightmost particle, then, is in the $-$ state, and faces the two $+$ particles on its left. The two outer particles (namely $1$ and $2$), whose interaction is screened by the middle particle, are pointing against each other. Their two-body potential, then, is higher than if they were parallel, so that the reduction in the three-body potential is greater than in the ${+}{+}{+}$ sector.
Second, the well at $d_{12}=d_{13}=1$ (\autoref{fig:RTP-HiLo}, left panel, bottom-left and top-right corners) is shallower than for $\Delta_{+++}$. The outer particles of this configuration ($1$ and $3$) are indeed aligned outwards, so that their two-body potential is minimal. Following this argument, it is natural that at $d_{12}=1,d_{23}=L-2$, where the two outer particles ($3$ and $2$) are aligned with each other, $\Delta_{-++}$ is similar to $\Delta_{+++}$.

Upon increasing $s$, the three-RTP potential of the ${+}{+}{+}$ sector becomes closer and closer to a pairwise potential in analogy with the TASEP potential (see the bottom-left corners of the colour plots of \autoref{fig:RTP-HiLo}). Conversely, in the  ${-}{+}{+}$ sector, the difference between three-body and pairwise potential increases with $s$. This observation holds for $\gamma>1$, i.e., in general, $\gamma >\omega$. For $\gamma \ll \omega$, i.e. approaching the limit of a symmetric simple exclusion process (SSEP), factorisation is achieved in both the  ${+}{+}{+}$ and the  ${-}{+}{+}$ sectors.

\section{Discussion}\label{conclusions}
%1. Smart int
%\begin{itemize}
%	\item Conditioning reveals smart interactions
%\end{itemize}
%2. 2-body results
%\begin{itemize}
%	\item TASEP not greater efficiency
%	\item RTP does at high Pe (related to time-scale)
%\end{itemize}
%3. general framework
%\begin{itemize}
%	\item Formula $\eta'$
%	\item Conditioning better close o PT (high var)
%	\item Limitation flat rf
%	\item formal construction (v1)
%\end{itemize}
%4.Towards many body
%\begin{itemize}
%	\item Approxiamtion, Whitelam, Escamilla
%\end{itemize}

Is conditioning a route to `smart' matter? The simplest example of just two interacting particles demonstrates that smart interaction can indeed emerge in this way: run-and-tumble particles develop an effective alignment interaction in order to sustain atypically large mobilities. This result provides a microscopic basis to the observation of aligned states in large work fluctuations of two-dimensional active Brownian particle systems~\cite{nemoto2019aa}. It also points towards a generality which extends beyond the one-dimensional continuous-time processes considered in this paper.
To judge whether conditioning yields an actual improvement on the individual energetics, we have proposed an efficiency framework which takes into account the energy-consuming nature of forces in active systems. Additionally, we have discussed the relationship between the effective potential in a two- and three-body scenario, which serves as a prototype for the generalization to higher particle counts.
%To judge whether the framework of conditioning holds any promise of yielding realistic interactions, we considered the energetic efficiency of the interactions, and the relationships between the effective potential in a two- compared to a three-body scenario as a prototype for the generalization to higher particle counts.

In terms of the efficiency, we discover in both the RTP and TASEP models that there is a phenomenon of rapidly diminishing returns, such that a relatively small window of conditioning values accounts for most of the range of possible efficiency gain. Furthermore, the relative amount of gain in efficiency differs substantially between the models. Conditioning can only `act' on naturally occurring fluctuations of the original dynamics, which are limited for the TASEP to fluctuations in hopping speed fortuitously correlated with inter-particle distances. In contrast, the RTP model, whose initial efficiency is lower due to head-to-head jamming, displays a broader repertoire of fluctuations to be exploited by conditioning (both speed and direction), which explains the larger efficiency gain compared to the TASEP.
Formula~\autoref{eq:etaprime} encapsulates this finding by providing a quantitative basis for the claim that a large variance in the output results in high gains in efficiency upon conditioning. In simple terms, when there are large and relatively likely fluctuations, conditioning can exploit them. At a mathematical level, high variance in output is equivalent to near-flatness of the saddle-point $s(\sigma)$. In a sense, this amounts to being close to a dynamical phase transition, at which the saddle-point would become truly flat, signalling the break-down of the large-deviation principle. However, as that happens, the ensemble equivalence that underlies the effective process construction is moot. We stress that studying the so-called $s$-ensemble, as is common, without relating it back to the value of the constraint $\sigma$, misses a qualitatively important aspect of conditioning, namely how the structure of the rate function itself determines the outcome of conditioning.

By comparing the two- and three-particle scenarios, we confirm that, in the general case, many-body interactions emerge that are not simple to extrapolate from the knowledge of the two-body interactions. While this may be perceived as a fundamental limitation of the conditioning approach, our detailed study of the three-body cases demonstrates that these many-body interactions need not be overly complicated. In the cases we examine, for instance, they can be ascribed to a topological screening effect: by placing an intermediate particle between two nearby ones, they are effectively screened, making the pairwise interaction across the intermediate particle superfluous. Thus, the 1D setup may be a main contributive factor to the lack of factorisation of the interaction. However, there are certainly situations in which factorization of the many-body interaction does occur, as in the high-current TASEP phase. To this we add the observation that, in the SSEP-limit of the RTP, the effective interaction factorises for arbitrary $s$. Conversely, for large \peclet number, three-body contributions to the RTP potential increase rather than decrease with the bias $s$. Future research may investigate more systematically what aspects of the dynamics lead to factorization (e.g., integrability and/or reversibility) while giving a thorough characterisation of three-body contributions when factorization is not expected.

Let us also stress that the conditioning framework is not limited to the arena of statistical mechanics. One may think of diverse practical scenarios where a specific potential or force is sought to achieve some outcome---this is the subject of optimal control theory, with which the concepts here discussed have been rigorously linked (see ~\cite{chetrite2015conditioning} and references therein). As in active and driven systems, it may be desired that the chosen constraint be satisfied only by adding a potential-like interaction, as the `tilt' factor $e^{s\alpha}$ implies an increased energy injection. To fix, in our language, the base hopping rate $\gamma$, one could consider a replica ($R$) of the naive process with hopping rate $\gamma_R \leq \gamma$ and choose a conditioning value $\sigma$ such that $R$ when conditioned on it attains a renormalized hopping rate $\gamma_R e^{s_{\gamma_0}(\sigma)} = \gamma$, i.e. $\sigma = s^{-1}_{\gamma_R}(\log(\gamma/\gamma_R))$. Finally, take $\gamma_R = \gamma^*$ as the value that optimizes the efficiency $\sigma/\gamma$ of this effective process,
\begin{equation}\label{eq:optsmart}
\gamma^* = \argmax_{\gamma_R} \left\{  s_{\gamma_R}^{-1}(\log \frac{\gamma}{\gamma_R}) \right\}.
\end{equation}
Construct the potential \autoref{eq:Veff} from the tilt of the transition matrix with hopping rate $\gamma^*$ and with tilt parameter $s^* = \log(\gamma/\gamma^*)$. This interaction potential added to the rates of the naive process with hopping rate $\gamma$ will have a higher efficiency (or at least not lower) while keeping energy input fixed. The price paid is that the resulting effective process is not strictly speaking representing the most probable fluctuations of the naive process it is compared to.

In closing, conditioning remains an intriguing framework to derive non-trivial interactions. It is intimately linked to inference from data \cite{Bialek2012,Cavagna2014}, and to optimal control. In the active matter context, the way conditioning exploits beneficial fluctuation is suggestive of an evolutionary point-of-view \cite{nemoto2019aa}, furthered by the similarity shared by rare-events sampling techniques~\cite{giardina2006aa} and gene selection. We have here only taken elementary steps in setting out the main ideas behind the framework and applying it to toy models. Our results nonetheless point to the effective process having a certain structure and robustness that generalizes with larger system sizes, provided the system parameters and conditioning value are chosen in a physically plausible way. While presently calculating large deviation elements of large systems is prohibitively costly, we expect concurrent developments of advanced approximations~\cite{Escamilla2019} and numerical methods~\cite{Jacobson2019x,Whitelam2018b} to overcome this hurdle.

\begin{acknowledgments}
FC acknowledges studentship support from SFC; EM from EPSRC grant no.\ EP/N509644/1.
FC and EM contributed equally to this work. The authors thank M.\ R.\ Evans for valuable feedback on the manuscript.
\end{acknowledgments}

\bibliographystyle{apsrev4-1}
%\bibliography{Refs}

\end{document}